\documentstyle[prl,aps,epsfig,rotate,epsf]{revtex}

\begin{document} 

\twocolumn[
\hsize\textwidth\columnwidth\hsize\csname@twocolumnfalse\endcsname
%\draft  

\title {
 $4D$ Spin Glasses in Magnetic Field Have a Mean Field like Phase}

\author{E. Marinari, F. Zuliani}
\address{Dipartimento di Fisica and INFN, Universit\`a di Cagliari\\
Via Ospedale 72, 09100 Cagliari (Italy)}

\author{G. Parisi}
\address{Dipartimento di Fisica and INFN, 
Universit\`a di Roma {\em La Sapienza}\\
P. A. Moro 2, 00185 Roma (Italy)}

\date{\today}

\maketitle

%\widetext
 
\begin{abstract}
By using numerical simulations we show that the $4D$ $J=\pm 1$ Edwards 
Anderson spin glass in magnetic field undergoes a mean field like 
phase transition.  We use a dynamical approach: we simulate large 
lattices (of volume $V$) and work out the behavior of the system in 
limit where both $t$ and $V$ go to infinity, but where the limit $V 
\to \infty$ is taken first.  By showing that the dynamic overlap $q$ 
converges to a value smaller than the static one we exhibit replica 
symmetry breaking.  The critical exponents are compatible with the 
ones obtained by mean field computations.
\end{abstract} 

\vfill
\twocolumn
\vskip.5pc ] 
\narrowtext

The mean field solution of spin glass systems \cite{PARBOO} contains 
many new features.  It tells us that systems with quenched disorder 
can have a large number of stable states, not related by explicit 
symmetries of the original Hamiltonian, and that the space of these 
states is embedded with an ultrametric structure.  Moreover, the 
system stays critical for all $T<T_{c}$ and the phase transition of 
Replica Symmetry Breaking (RSB) survives the presence of a finite 
magnetic field $h$.

The mean field paradigm needs to be analyzed, in order to understand 
how many of its peculiar features are shared by the finite 
dimensional, physically relevant case.  In spite of the technical 
difficulties, in the last years many progresses have been done.  It is 
for example remarkable that recent rigorous results \cite{RIGORA} seem 
to support strongly (after some initial different feelings 
\cite{RIGORB}) the viability of the mean field approach for the 
description of finite dimensional systems. It has been shown 
\cite{RIGORA} that the rigorous finite dimensional construction of 
\cite{RIGORB} leads to self-averaging quantities exactly where the 
mean field construction would also produce self-averaging observables, 
and Guerra has shown that the main part (and maybe all) of the replica 
predictions on the fluctuations of non-self-averaging quantities 
applies to the broken phase  of finite dimensional disordered systems 
(see also \cite{TRE}).

Monte Carlo simulations are an important tool to establish how much of 
the mean field description survives in the finite dimensional case 
\cite{MCR,MC}.  For example there is now evidence for the existence of 
a $3D$ mean-field like critical point, for the existence of an 
ultrametric structure in $4D$, and for a dynamical behavior of finite 
dimensional systems very similar to the one that can be found 
analytically in the Sherrington-Kirkpatrick mean field model.

The question of the existence of a de Alme\-ida-Thou\-less \cite{DAT0} 
line, i.e.  of the existence of a phase transition in finite magnetic 
field, is maybe the most relevant open problem.  Even if a large 
amount of numerical work has been done to clarify this issue 
\cite{DAT1,DAT2}, a clear cut answer is still lacking.  Most of the 
numerical work suggests that a transition exists (even if some studies 
suggest the opposite conclusion), but the question is a very delicate 
one: one finds probability distributions that do not have a very clear 
behavior, and it is very difficult to thermalize large systems in the 
low temperature ($T$) region.  Even the most recent numerical work of 
\cite{DAT2} does not reach unambiguous conclusions.

Here we hope to settle the question, by showing in a non-ambiguous way 
that the $4D$ spin glass with quenched couplings $J=\pm 1$ in finite 
magnetic field undergoes a mean field like phase transition.

We use a dynamical approach.  If a large system is cooled down to a 
temperature $T_{f}$, starting from the high temperature region, after 
a time $t$ the correlation functions are different from zero (in a 
statistically significant way) only up to distances smaller than a 
dynamic correlation length $\xi(t)$.  Often (and this seems to be the 
case of spin glasses in the low $T$ phase) $\xi(t)$ increases as a 
power of $t$, i.e.  $\xi(t) \propto t^{1/z(T)}$.  If the lattice size 
$L$ is larger than $\xi(t)$, for large times the system is locally, 
but not globally thermalized: in the case of an infinite lattice this 
is always the case, independently from the value of $t$.  As we shall 
see later our choice of the lattice volume, $V=20^{4}$, is such that 
we stay in this situation.  Then by using power fits we keep the large 
time limit under control, and we determine with high precision the 
infinite time expectation values, always in the phase where 
$L>\xi(t)$.

A key prediction of the theory of replica symmetry breaking is that 
when comparing different realizations of the system we find that there 
are local quantities which take a different value in the $t\to\infty$ 
and $V\to\infty$ limits in the two regions $L \gg \xi(t)$ and $L \ll 
\xi(t)$.  In the following we will call respectively dynamic and 
static the expectation values computed in the first and in the second 
region.  We aim to show that in four dimensional spin glasses in 
magnetic field at low temperature the two expectation values are 
different and therefore the replica symmetry is broken, as expected 
from mean field computations.

This work contains two kind of results.  First of all we discuss some 
inequalities, both at finite $h$ and in the $h\to 0$ limit, that can 
be violated only if replica symmetry is broken.  We use our numerical 
simulations to show that indeed such inequalities are broken for $T$ 
small enough.  Second we show that our data for the overlap and for 
the underlying time scales obey an impressive scaling versus the 
magnetic field, and that the critical exponents turn out to be very 
similar to the mean field theoretical prediction.

Numerical data are drawn from dynamical runs scheduled according to 
the following scheme.  We start at $T_{1}=3.00$ (the value of the 
critical temperature at $h=0$ is close to $2.0$) and decrease $T$ with 
steps of $0.25$ down to $T_{9}=1.00$.  For an annealing run of level 
$k$ at $T=T_{n}$, $n\cdot 2^{k}$ steps are performed with $k$ spanning 
from $5$ to $16$ ($14$ for the larger fields) and $h$ from $0.2$ to 
$0.6$ (runs at $h=0.1$ do not reach enough precision to be used for 
fitting, and have only been included in the matching analysis, see 
later).  In total our longer runs involve order of $3$ millions sweeps 
of the $20^{4}$ lattice.  The quantity $t\equiv 2^{k}$ plays the role 
of a time: we will be extrapolating expectation values on $k$ at given 
$T$ and $h$.  We use two copies of the system $\sigma$ and $\tau$ in 
each realization of the quenched couplings to compute the overlap $q 
\equiv \sum_{i}\sigma_{i}\tau_{i}/V$.  We use a multispin coded 
algorithm \cite{ZULLO} that allows to flip more than $70\cdot 10^{6}$ 
spins per second on a Digital $\alpha$ workstation $500/333$.  We have 
averaged over $64$ samples for each $k$ and $h$ value (for a very few 
cases we only have $32$ samples).  In principle we could also put the 
system at the final temperature by a a sudden quench.  We have 
followed the previous procedure for two reasons: (a) Finite time 
effects are smaller and the infinite time extrapolation is easier.  
(b) We can collect in one run data at different temperatures.

The first kind of evidence is based on our results for $q(t)$ at fixed 
$h$. We extrapolate $q(t)$ to its value for infinite time $q^{D}$ 
($q^D$, where $D$ stands for dynamical, is also the minimum allowed 
value for $q$ at equilibrium, $q_{min}$ \cite{DAT1}).  We compare 
$q^D$ with the static value $q$ computed with equilibrium runs 
for a $7^4$ system \cite{DAT2} (preliminary runs of \cite{DAT3} 
confirm the determination of $q^S$ of \cite{DAT2} down to $T=1.00$).

\begin{figure}
  \epsfxsize=1.1\columnwidth\centerline{{\epsffile{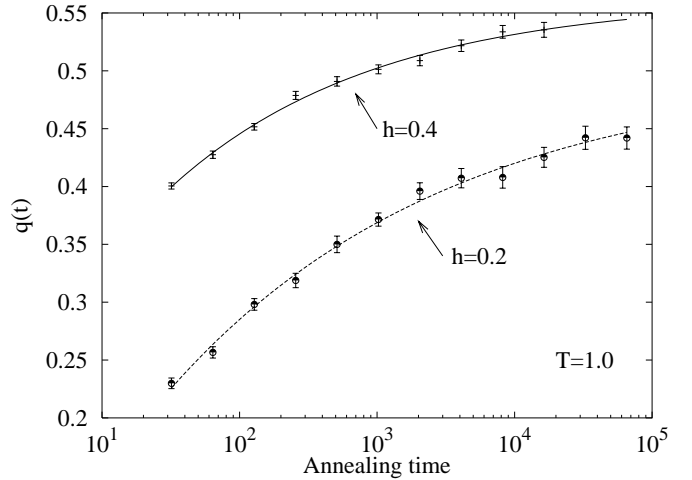}}}
  \protect\caption[1]{
    $q(t)$ versus $t$ and best fit.
  }
  \protect\label{F-ONE}
\end{figure}

We show that at low $T$ $q^{D} < q^{S}$ strictly, i.e.  that replica 
symmetry is broken.  In fig.  (\ref{F-ONE}) we show two typical fits 
for low $T=1.0$, at $h=0.4$ and $h=0.2$ (here we are using for fitting 
all the data points: see later for scaling time windows).  Moreover, 
since the values of $q$ increase with the lattice sizes in the static 
runs of \cite{DAT2}, our evidence is safe also from the point of view 
of finite size effects.

The power fits are very good.  One finds $q(t)$ $=$ $0.56(1)$ $+$ 
$0.44(3)$ $t^{-.28(3)}$ and $q(t)$ $=$ $0.50(2)$ $+$ $0.58(2)$ 
$t^{-.21(3)}$ respectively at $h=0.4$ and $0.2$.  Both fits have a 
very good $\chi^{2}$.

\begin{figure}
  \epsfxsize=1.1\columnwidth\centerline{{\epsffile{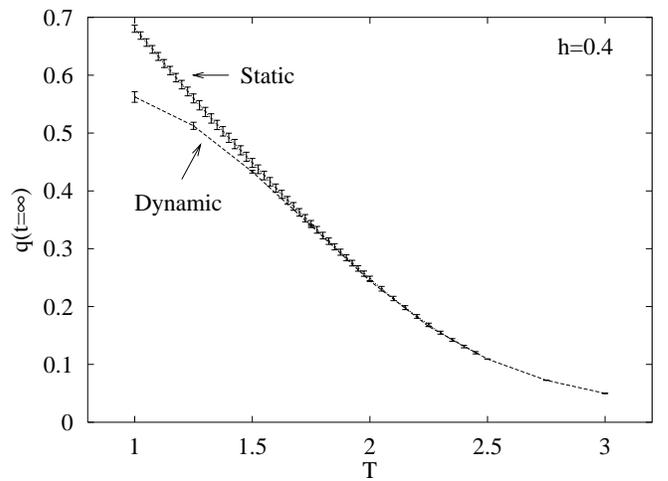}}}
  \protect\caption[1]{
    $q^{D}$ from our runs and $q^{S}$ from \protect\cite{DAT2} 
    versus $T$.
  }
  \protect\label{F-TWO}
\end{figure}

In figure (\ref{F-TWO}) we plot our data for $q^{(D)}$ (dashed 
curve and error bars) at $h=0.4$ and the static data of 
\cite{DAT2}.  For high values of $T$ the data are in perfect 
agreement, while at $T=1.5$ the two curves start to split in a 
statistically significant way.

Both at $T=1.25$ and at $T=1.0$ it is clear that the difference of 
the dynamic and the static value is both statistically and 
systematically significant with a large confidence level.  So at 
$h=0.4$ we have evidence that for $T\le 1.25$ the system is in a 
mean field like broken phase.  These results are in good agreement 
with the data of \cite {DAT2} that suggests a transition near 
$T=1.5$ at this value of the magnetic field.

The reader could wonder if in our simulations the inequality 
$L\gg\xi(t)$ is satisfied.  By estimating the exponent $z(T)$ from the 
simulations at $h=0$ we find that $z \approx 10$ at $T=1.0$ and that 
the bound should be saturated for times $O(10^{13})$, which is much 
larger than the largest times scales of our numerical simulation.  
Simple power fits to energy, overlap and magnetization and to their 
fluctuations are good: this is an independent indication of the fact 
that that all our data are in the region where $L\gg \xi(t)$.  
Moreover, even if $L$ was close to $\xi(t)$ our conclusion would be 
strengthened since then the difference among the measured dynamical 
value and the static value could only decrease.

Next we discuss the $h\to 0$ limit.  We consider the susceptibility 
$\chi\equiv \lim_{h\to 0}\frac{m}{h}$.  If replica symmetry is 
realized we have that for $h\to 0$ the susceptibility $\chi =\beta 
(1-q)$, where there is no ambiguity in the definition of $q$.  We can 
thus define

\begin{equation}
\tilde q(h) \equiv 1 -T \frac{m(h)}{h}\ .
\end{equation}

\begin{figure}
  \epsfxsize=1.1\columnwidth\centerline{{\epsffile{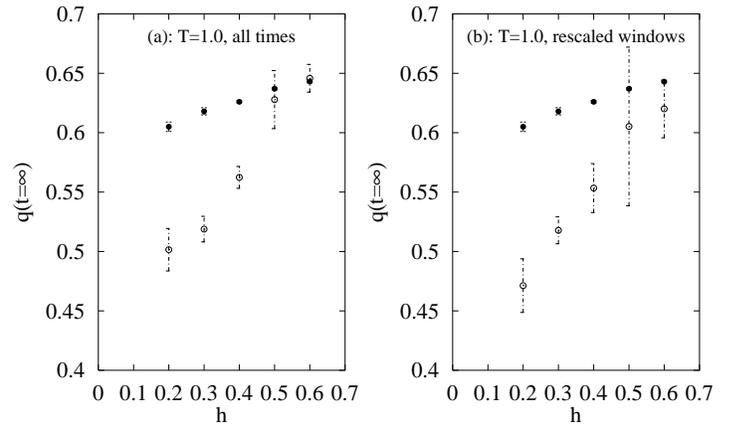}}}
  \protect\caption[1]{ 
    $q^{D}$ and $\tilde{q}$ from fits including all 
    time points and from fits done over rescaled time windows.} 
  \protect\label{F-THREE}
\end{figure}

If in the $h\to 0$ limit $\tilde q\ne q^{D}$, than replica symmetry is 
broken.  In a theory where replica symmetry is broken the small $h$ 
limit of $\tilde q$ is $q^{S}$.  More precisely for finite $h$ we find 
that $\tilde q(h)=q^{S}(h)+O(h^{2})$.  In fig.(\ref{F-THREE}.a) we 
show $q^{D}$ at $T=1.0$ as a function of $h$ (empty dots), together 
with the values of $\tilde q(h)$ (filled dots).  The two functions do 
not extrapolate to the same value at $h=0$.  Replica symmetry is 
broken in the region where $q^{D}$ is smaller than the $h=0$ limit of 
$\tilde q$.  If we neglect terms of order $h^{2}$ (the difference 
among $\tilde q$ and $q^{S}$ is about .02 at $h=.4$), replica symmetry 
must be broken when the two curves differs in a statistically 
significant way (i.e., in our case at $T=1.0$, for $h<0.5$).

As we will discuss later we have determined a rescaling of times as a 
function of $h$ that makes the curves $q(t)$ at different fields 
universal.  We can thus determine consistent $h$ dependent time 
windows, that allow us to compare homogeneous time regimes at the 
different $h$ values.  The $k$-windows used for these scaling fits 
($t\equiv 2^{k}$) are $8-16$ at $h=0.2$, $7-14$ at $h=0.3$, $6-13$ at 
$h=0.4$, $5-11$ at $h=0.5$ and $5-10$ at $h=0.6$.  The results of the 
corresponding fits are given in figure (\ref{F-THREE}.b).  Here the 
points have somehow a larger error (since we use less data point for 
fitting) but we expect the systematic error to be smaller.  The points 
at $h=0.2$, for example, appear more consistent thanks to the 
elimination of short time effects.  The emerging physical picture is 
independent from the fitting scheme.

For small $h$ in the SK model $\tilde q(h)=\overline{q}(0)+ R
q_{D}^{2}$, where $R\simeq P(0)/2$, and $P(0)$ is the value of the
function $P(q)$ at $q=0$ when $h=0$.  We have found that in the region where
$\tilde q \ne q^{D}$ the data are compatible with a quadratical
dependence over $q^{D}$.  We find for example $R=.4$ at $T=1.25$,
which is of the same order of magnitude of $P(0)/2$
(the value of $P(0)$ at this temperature is about $.5$ \cite{DAT3}).

Let us give some information about the exponent of the power-law fit 
we have determined for the decay of $q(t)$.  At low $T$ ($1.0$ and 
$1.25$) such exponents are between $.2$ and $.3$ with no apparent 
systematic dependence on the magnetic field.  The fits on rescaled 
time windows give higher values than the fits on all points (basically 
for low $T$ values the results are fixed around $\frac{1}{3}$).  We 
have also fitted the energy with a power decay to its asymptotic 
value.  Here the decay exponent can be estimated with good precision, 
and for low $T$ it does not depend on $h$.  For example at $T=1.00$ we 
find an exponent of $.435$ for $h$ going from $0.1$ to $0.4$.  At 
$T=1.25$ it is $0.47$ for $h$ going from $0.1$ to $0.3$, while at 
$T=1.5$ it already has a small dependence on $h$.

Our last numerical evidence is based on rescaling the functions 
$q_{h}(t)$ obtained for different values of $h$. We have rescaled our 
data obtained at different magnetic field values  according to
$q_{h'}(t') = A(h) q_{h}(B(h) t)$.

The coefficients for the rescaling to a fixed value of $h'$ are fitted 
to the form $A(h)\simeq \tilde{A} h^{\omega}$, and $B(h)\simeq 
\tilde{B} h^{\tau}$.

The value of the crossover exponents may be found by assuming that the 
quantity $\int d^{4}x \ h(x)^{2}q(x)$ is dimensionless: the coupling 
term in replica space has a form $\int d^{4}x \ 
h^{2}(x)\sum_{a,b}Q_{a,b}(x)$.  The dimension of $q$ in the dynamic 
approach can be reconstructed by the decay of the correlation 
functions.  An approximate formula (which seems to work reasonably 
well~\cite{QUATTRO}) has been proposed in~\cite{DKT}: $\langle 
q_{x}q_{0}\rangle \propto x^{-\lambda}$, with $\lambda \approx 
(D-2)/2$.  This formula, together with dimensional analysis, implies 
that in $D=4$ $\omega\simeq\frac27$.  A similar analysis shows that in 
$D=3$ $\lambda= \frac{2}{11}$.  Since the dynamical critical exponent 
is of order $8$ in the temperature range around $T=1.25$ the same 
argument implies that $\tau\simeq 4$. 

We have determined the best coefficients $A$ and $B$ by minimizing the 
square difference of the two functions.  Below we discuss results at 
$T=1.0$.  The fits are very good: all the rescaled $q$ have ratios 
systematically compatible with $1$, and there is no need for a further 
extrapolation or corrections to our scaling formula.  

\begin{figure}
  \epsfxsize=250pt\centerline{{\epsffile{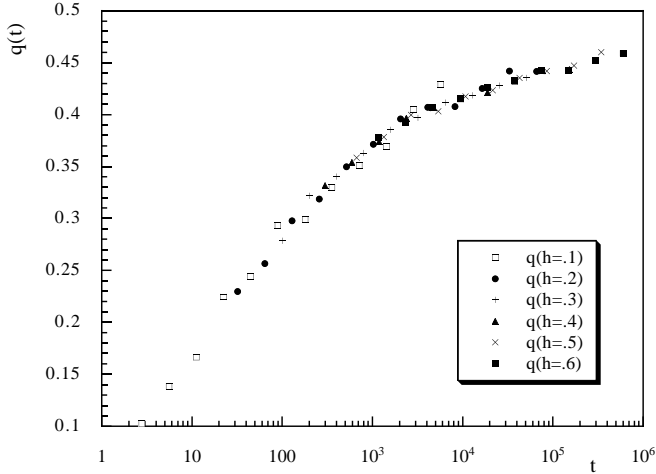}}}
  \protect\caption[1]{
    Rescaled functions $q(t)$ versus annealing time.
  }
  \protect\label{F-FIVE}
\end{figure}

\begin{figure}
  \epsfxsize=1.0\columnwidth\centerline{{\epsffile{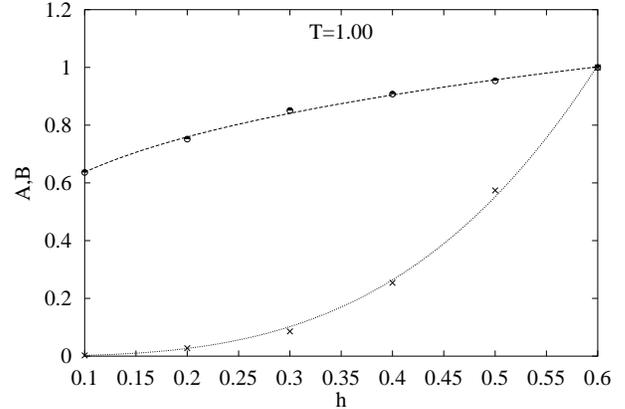}}}
  \protect\caption[1]{
    $A$ and $B$ from our best fit versus $h$.
  }
  \protect\label{F-FOUR}
\end{figure}

In figure (\ref{F-FIVE}) we show the rescaled functions (horizontal 
and vertical scales are given by the fact that we have kept fixed the 
values at $h=0.2$).  The scaling is obeyed remarkably well: it works 
over six time decades, and in a range of magnetic fields going from 
$.1$ to $.6$.  The errors on data point are not plotted since they 
would blur the figure.  They are of the order of $0.01$-$0.02$.  For 
example the point at $h=0.1$ with largest $t$ values, that is slightly 
out of the enveloping curve, is statistically compatible with the 
other points.  $A$ and $B$ determined with the fits of figure 
(\ref{F-FIVE}) can now be fitted with power laws.  We plot them in 
figure (\ref{F-FOUR}), with $A$ represented by the upper points and 
$B$ from the lower ones, together with the best fits (the fitting 
function are normalized in such a way to give $1$ at $h=0.6$).  The 
best fit gives $A(h)\simeq 1.14(1) h^{.25(1)}$, and $B(h)\simeq 5.4(3) 
h^{3.3(1)}$.  Even if this is a qualitative test, since we have only a 
rough estimate from the mean field approach, and in this case we have 
not analyzed the statistical and systematic error in great detail 
(since systematic error could be quite large for this measurement) the 
agreement with the values one would expect from the mean field 
solution turns out to be remarkably good.

We acknowledge useful discussions with C. Naitza, F. Ricci-Tersenghi 
and J. J. Ruiz-Lorenzo.  We warmly thank F. Ritort for interesting 
correspondence and comments.

%%%%%%%%%%%%%%%%%%%%%%%%%%%%%%%%%%%%%%%%%%%%%%%%%%%%%%%%

\end{document}